\documentclass[a4paper, 10pt]{article}
\usepackage{rits2013} % Fichier style pour RITS 2013

\usepackage[latin1]{inputenc} % Utilisation des accents 
\usepackage[french]{babel} % Mots réservés français
\usepackage{float}
\usepackage{graphicx}

\title{Lorentz Force Electrical Impedance Tomography}

\name{Pol Grasland-Mongrain$^1$, Jean-Martial Mari$^1$, Jean-Yves Chapelon$^1$, Cyril Lafon$^1$}

\address{
$1.$\, Inserm, U1032, LabTau, Lyon, F-69003, France ; Université de Lyon, Lyon, F-69003, France
}

\begin{document}
\maketitle

\resume{
This article describes a method called Lorentz Force Electrical Impedance Tomography.
The electrical conductivity of biological tissues can be measured through their sonication in a magnetic field: the vibration of the tissues inside the field induces an electrical current by Lorentz force. This current, detected by electrodes placed around the sample, is proportional to the ultrasonic pressure, to the strength of the magnetic field and to the electrical conductivity gradient along the acoustic axis. By focusing at different places inside the sample, a map of the electrical conductivity gradient can be established. In this study experiments were conducted on a gelatin phantom and on a beef sample, successively placed in a 300 mT magnetic field and sonicated with an ultrasonic transducer focused at 21 cm emitting 500 kHz bursts. Although all interfaces are not visible, in this exploratory study a good correlation is observed between the electrical conductivity image and the ultrasonic image. This method offers an alternative to detecting pathologies invisible to standard ultrasonography.}
\motscles{Medical Imaging, Electrical Conductivity Imaging, Magneto-Acousto-Electrical Tomography, Ultrasonically-Induced Lorentz Force, Lorentz Force Electrical Impedance Tomography, Hall Effect Imaging}
\section{Introduction}
The electrical conductivity of biological tissues arouses great interest for medical imaging researchers. Indeed this parameter potentially shows a good contrast in the human body. For example, fat is ten times less conductive than muscle tissue \cite{gabriel1996} while the acoustic impedance, observed in ultrasonography , only changes of a few percents through the soft tissues. Electrical conductivity anomalies can moreover reveal pathologies like tumors \cite{haemmerich2003}.

The most advanced technique today to measure the electrical conductivity of tissue is the Electrical Impedance Tomography (EIT) \cite{cheney1999}. In this technique, several electrodes are placed around the body or the organ under study. An electrical current is injected through each electrode while its distribution in the tissues is measured by the others. These measurements allow an electrical impedance image reconstruction through mathematical analysis. This apparently simple and inexpensive technique suffers however of a low spatial resolution due to the intrinsic ill-posed nature of the mathematical problem \cite{ammari2004}.

On another hand, the vibration of a conductor inside a magnetic field induced by ultrasound induces by Lorentz force an electrical current which is connected to the electrical conductivity of the conductor \cite{filipczynski1969}. This approach can be applied to the measurement of ultrasound velocity using a wire as sensing element, for its electrical conductivity is known \cite{grasland2012},   \cite{grasland2013}. Conversely,the ultrasound pressure distribution in the conductor is known  \cite{mari2010approximate}, the electrical conductivity could be deduced . In this last approach, the biological tissues are submitted to a magnetic field created for example by a permanent magnet, and a focused ultrasound beam is used to vibrate the tissues in a specific region of interest \cite{mari2009}. In the same way the movement of a conductor in a magnetic field would induce an electrical current, this vibration induces a current in the tissues which can be detected by the mean of electrodes. The focusing of ultrasound allows conferring to the imaging process an important characteristic when compared to Electrical Impedance Tomography: the spatial resolution is close to the one that would be reached by ultrasound imaging. However, the drawback of such approach lies in the weakness of the induced electrical current. The initial name of the technique, the Hall Effect Imaging \cite{wen1998} was criticized later on as the phenomenon is not exactly an Hall Effect \cite{roth1998},. The technique has also been called Magneto-Acousto-Electrical Tomography \cite{haider2008} or scan of electric conductivity gradients with ultrasonically-induced Lorentz force \cite{montalibet2002scanning}. We use here the name of Lorentz Force Electrical Impedance Tomography (LFEIT), which has the advantage of describing both the imaged parameter and the method. 

\section{Theory}
For clarity reasons, the X axis is defined as the orientation of the magnetic field, the Z axis is defined along the ultrasound propagation direction and the Y axis is placed accordingly using the right-hand rule. As exposed previously, the Lorentz Force Electrical Impedance Tomography is based on the movement of a conductor placed inside a magnetic field, which induces an electrical current. Physical and mathematical modellings have been proposed to describe the method \cite{ammari2009}, \cite{roth1994}. We choose here the model presented by Montalibet et al. which gives a good understanding of the phenomenon \cite{montalibet2001electric}. In this model, the induced current is found to be proportional to the convolution product of the electrical conductivity gradient $H$ with the ultrasound pressure shape $P$:
\begin{equation}
	i(t) = \int{ (\frac{d\sigma }{dz} \frac{B}{\rho}) * (\int_0^t{p(\tau)d\tau} ) dS} = c(H*P)(t)
\end{equation}
with $i$ the induced electrical current, $\frac{d\sigma}{dz}$ the electrical conductivity gradient along z axis, $B$ the magnetic field, $\rho$ the density, $p$ the ultrasound pressure and $c$ the speed of sound.

\begin{figure}[t]
 \centering
 \includegraphics[width=0.9\columnwidth]{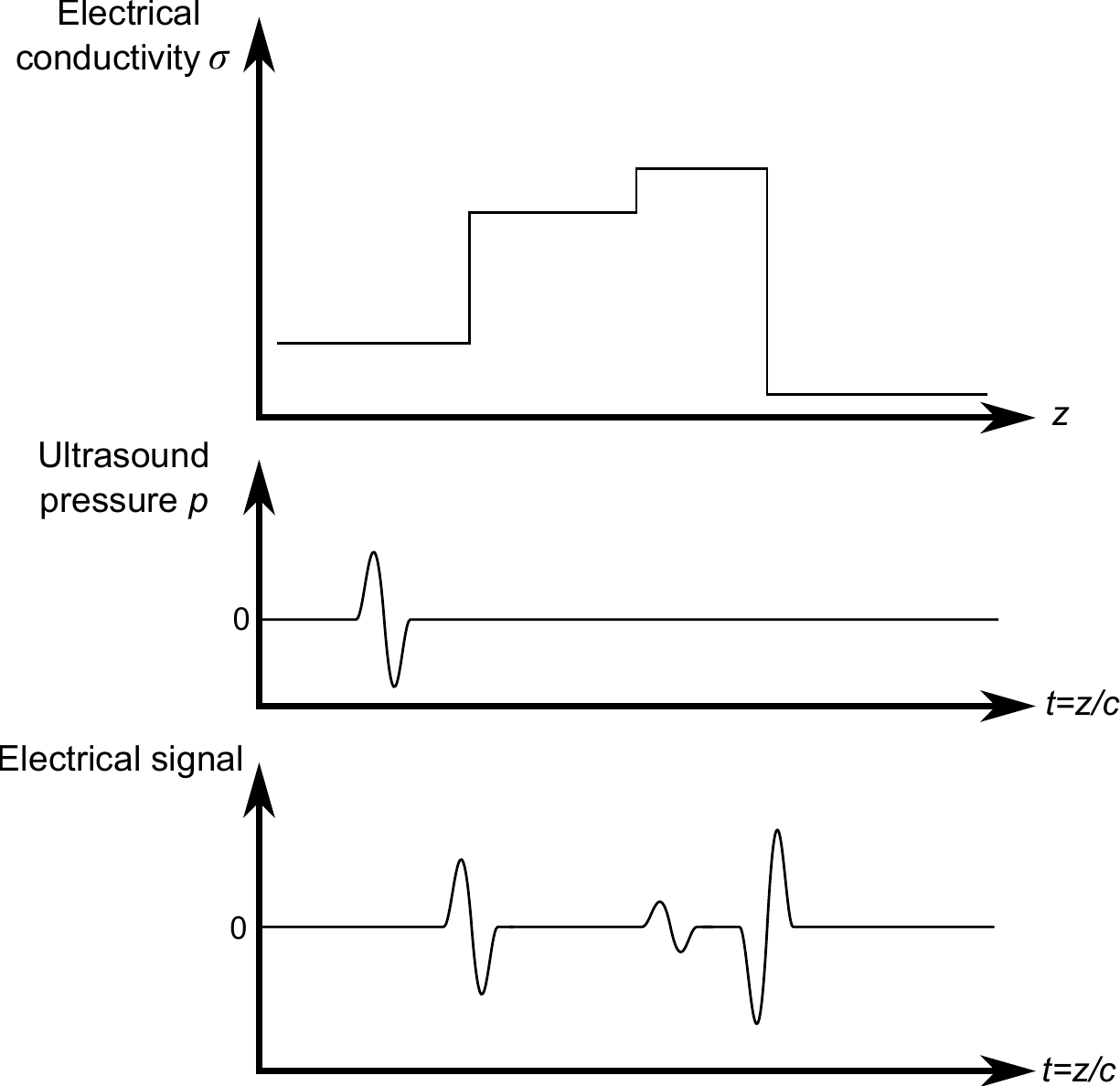}
 \caption{The electrical signal detected by electrodes in Lorentz Force Electrical Impedance Tomography is proportional to the convolution product between electrical conductivity gradients and the ultrasound pressure shape.}
 \label{theory}
\end{figure}
In other words, the electrical current detected by the electrodes is a temporal image of the ultrasound pulse at each electrical conductivity interface, as shown by the figure (\ref{theory}). More refined models are however available \cite{ammari2009}, \cite{roth1994}. 

\section{Materials and Methods}
The goal of this study was to build a setup to produces LFEIT images of gelatine phantoms and biological tissues.

\begin{figure}[t]
 \centering
 \includegraphics[width=0.9\columnwidth]{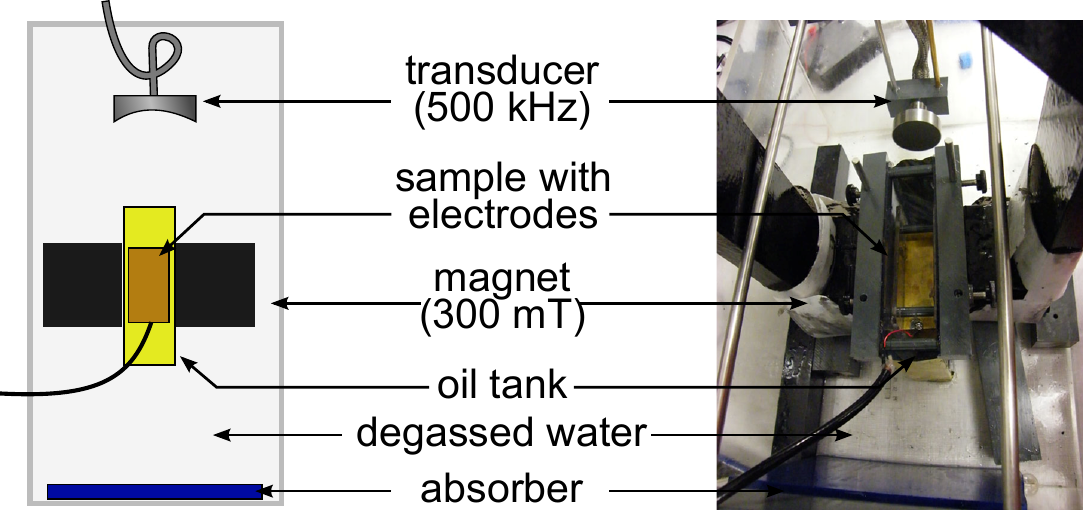}
 \caption{A transducer is emitting ultrasound in a sample placed in an oil tank in the middle of a magnetic field. The induced electrical current is received by two electrodes.}
 \label{schema}
\end{figure}

The experiment setup is illustrated in figure (\ref{schema}). A generator (HP33120A, Agilent, Santa Clara, CA, USA) created 0.5 MHz, 3 cycles sinusoid bursts at a pulse repetition frequency of 100 Hz. This excitation was amplified by a 200 W linear power amplifier (200W LA200H, Kalmus Engineering, Rock Hill, SC, USA) and sent to a 0.5 MHz, 50 mm in diameter transducer focused at 210 mm and placed in a 100x50x50 cm$^3$ degassed water tank.

\begin{figure}[t]
 \centering
 \includegraphics[width=0.98\columnwidth]{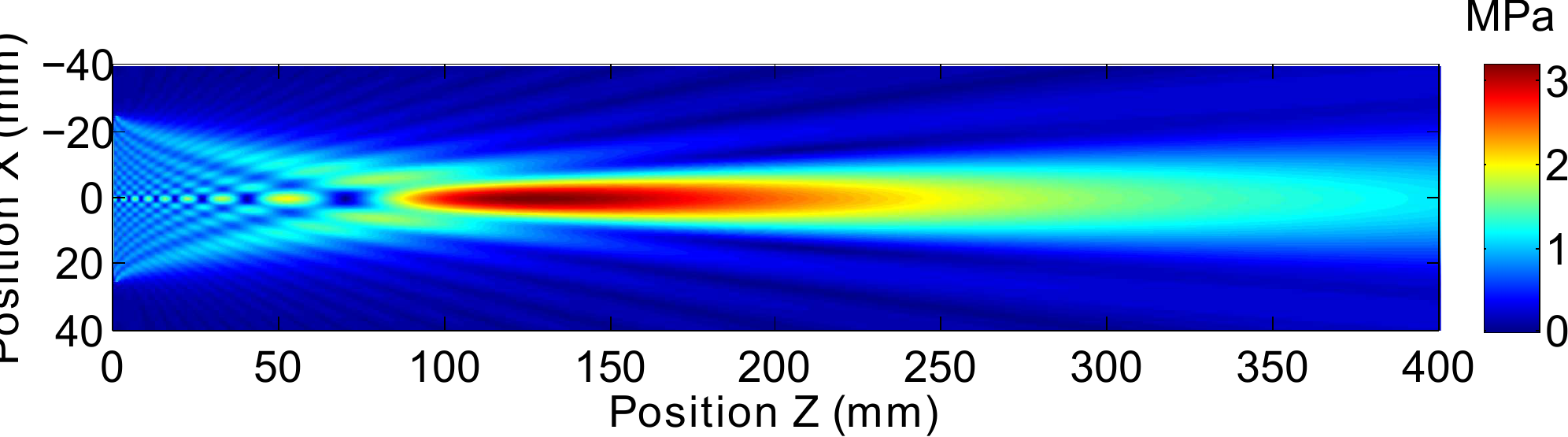}
 \caption{ Pressure field simulation along ultrasound axis ($Y$ = 0 mm) of the transducer. The maximum of pressure is approximately at 15 cm of the transducer.}
 \label{pressureField}
\end{figure}

The ultrasound pressure was equal to 3 MPa at the focal point (Z = 210 mm). A simulation of the ultrasound pressure field based on the resolution of the linear Rayleigh equation is shown in figure (\ref{pressureField}).

A 20x4x15 cm$^3$ mineral oil tank was placed from 15 to 35 cm away from the transducer in the ultrasound beam axis. This oil tank was placed in the middle of a U-shaped permanent magnet, composed of two poles made of two 5x5x3 cm3 NdFeB magnets (BLS Magnet, Villers la Montagne, France). The gap between the poles was 4.5 cm. The magnetic field was equal to 90 mT in a 4 cm radius circle around the maximum of 340 mT.

\begin{figure}[t]
 \centering
 \includegraphics[width=0.9\columnwidth]{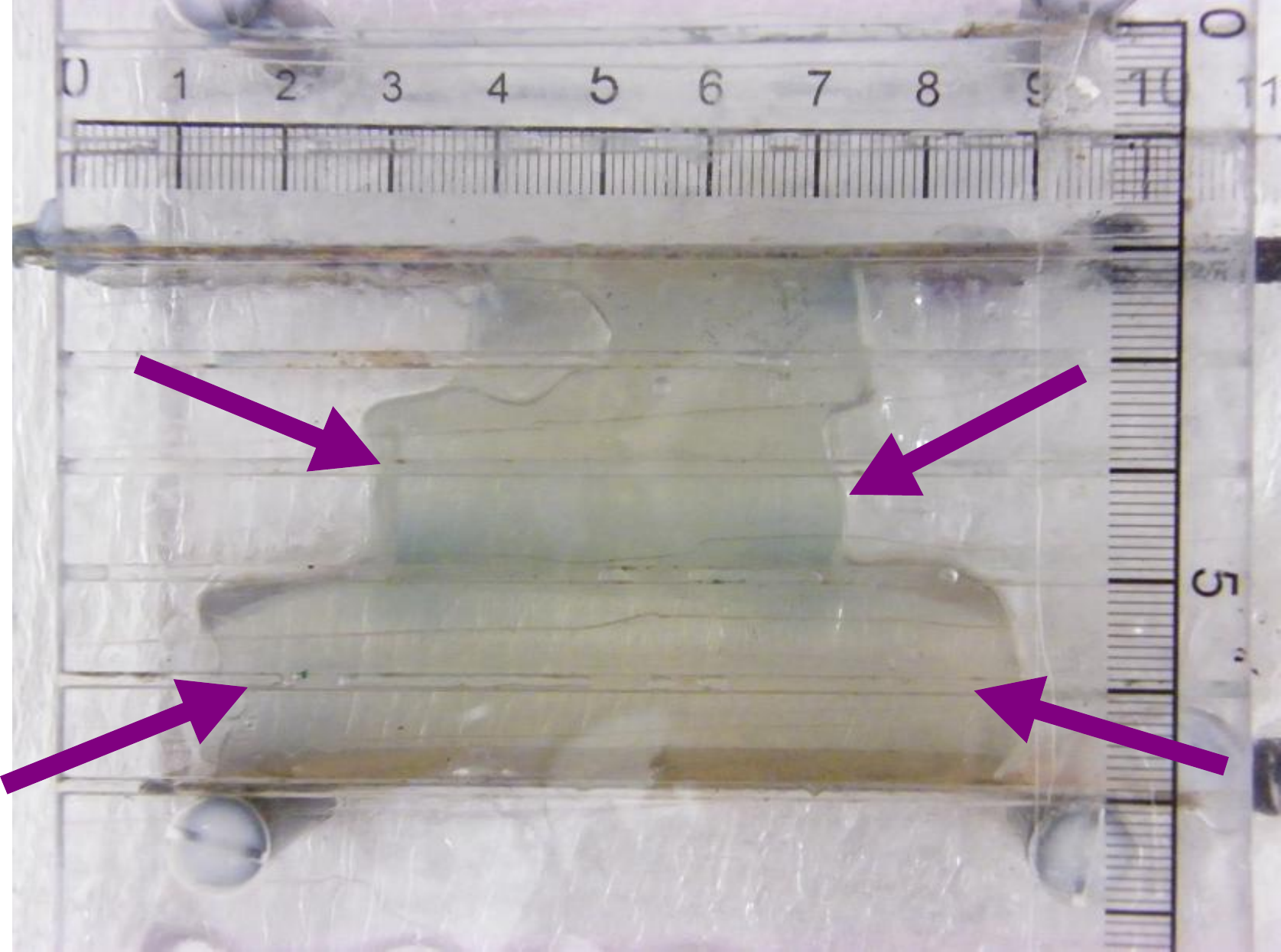}
 \caption{Photo of salty gelatin sample. The sample is made of 2 blocks, one of 8x2x2 cm$^3$ under the second of 4x2x3 cm$^3$. Two electrodes are in contact above and under the sample. Arrows are indicating front and back interfaces.}
 \label{photo_gelatin}
\end{figure}

\begin{figure}[t]
 \centering
 \includegraphics[width=0.9\columnwidth]{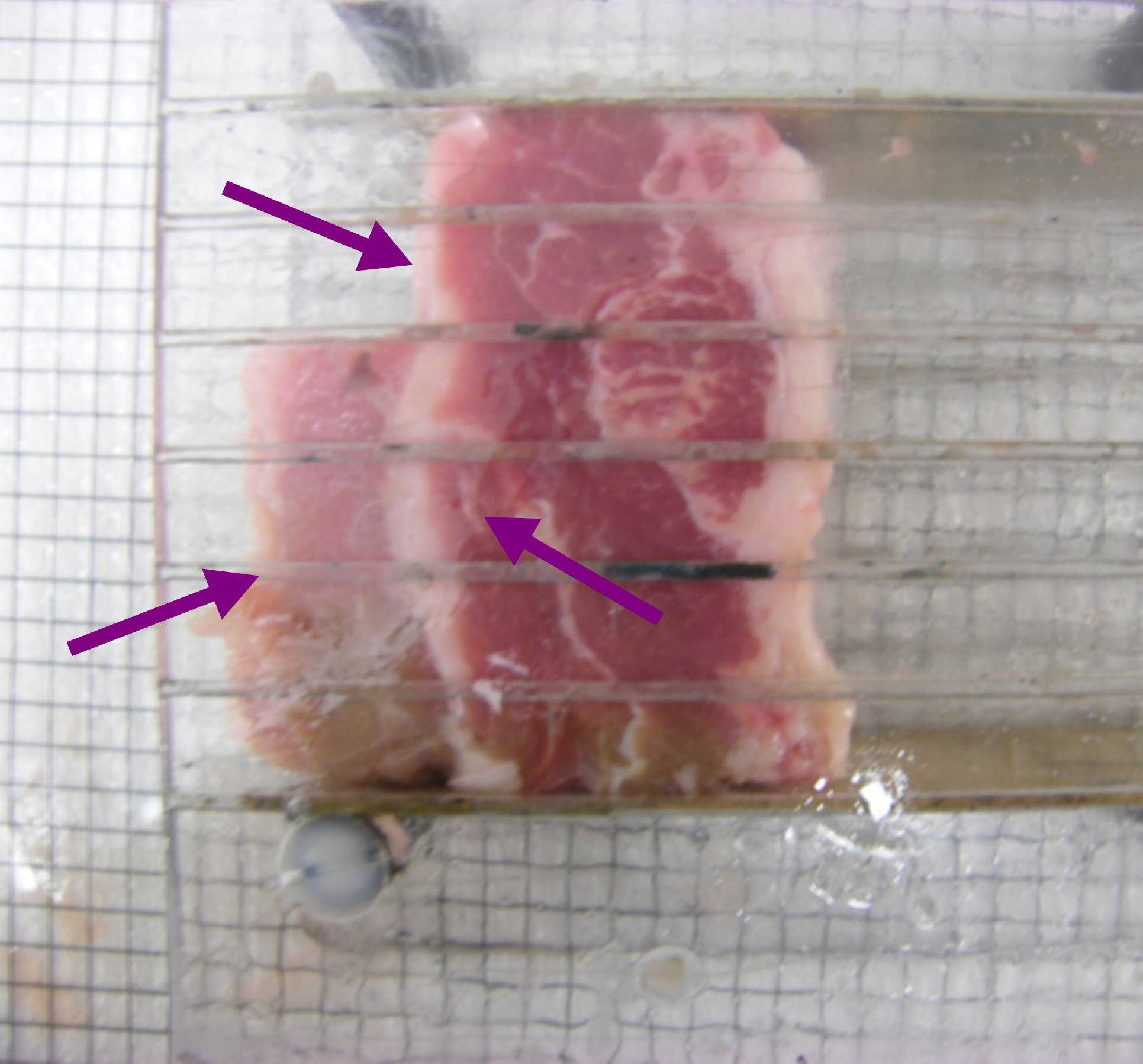}
 \caption{Photo of beef sample. The sample has an L-shape, 6 cm wide, 2 cm large and 6 cm high. Two electrodes are in contact above and under the sample. Squares are 5 mm wide. Arrows are indicating two front interfaces and the fat layer.}
 \label{photo_beef}
\end{figure}

The tested sample was placed inside this tank from 22 to 28 cm of the transducer. Two samples were tested: a phantom made of 10\% gelatin and 1\% salt, 8 cm wide, as shown in figure (\ref{photo_gelatin}); and a 6 cm wide piece of beef muscle with an L-shape, having a fat layer in the middle of the sample, as shown in figure (\ref{photo_beef}). A pair of 10x3x0.1 cm$^3$ copper electrodes was placed in contact with the sample, respectively above and under it. The electrodes were linked through an electrical wire to a 1 MV/A current amplifier (HCA-2M-1M, Laser Components, Olching, Germany) and an oscilloscope with 50 $\Omega$ input impedance (WaveSurfer 422, LeCroy, Chestnut Ridge, NY, USA) which was averaging the measures over 5000 acquisitions. The signals were post-processed using the Matlab software (The MathWorks, Natick, MA, USA) by computing the Hilbert transform modulus of the signal and converting it into a line of color.

The ultrasound signal was simultaneously recorded from the echoes on the transducer.

\section{Results}
\begin{figure}[ht]
		\begin{center}
	   		\includegraphics[width=.9\linewidth]{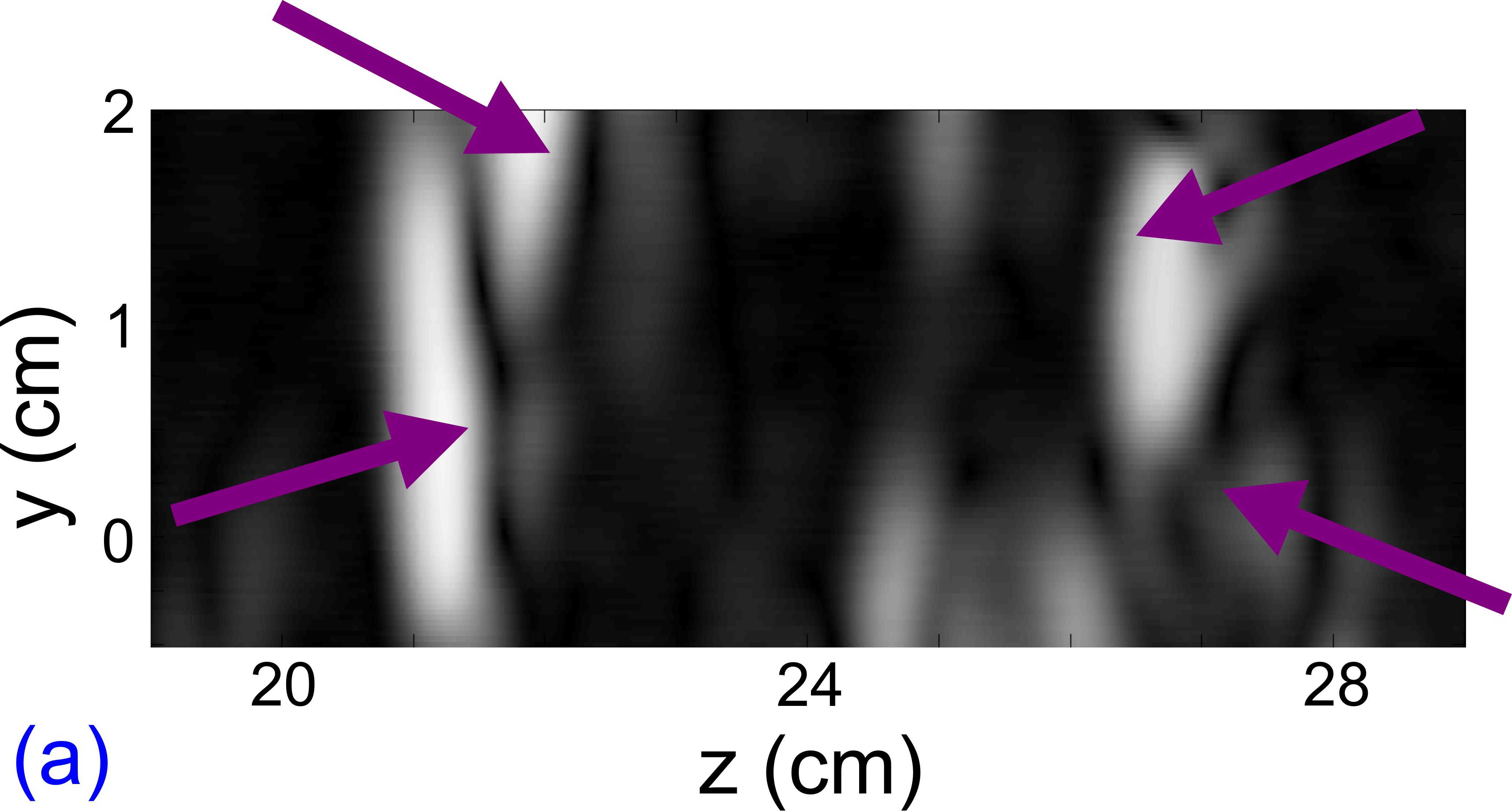}
		\end{center}
		\begin{center}
	   		\includegraphics[width=.9\linewidth]{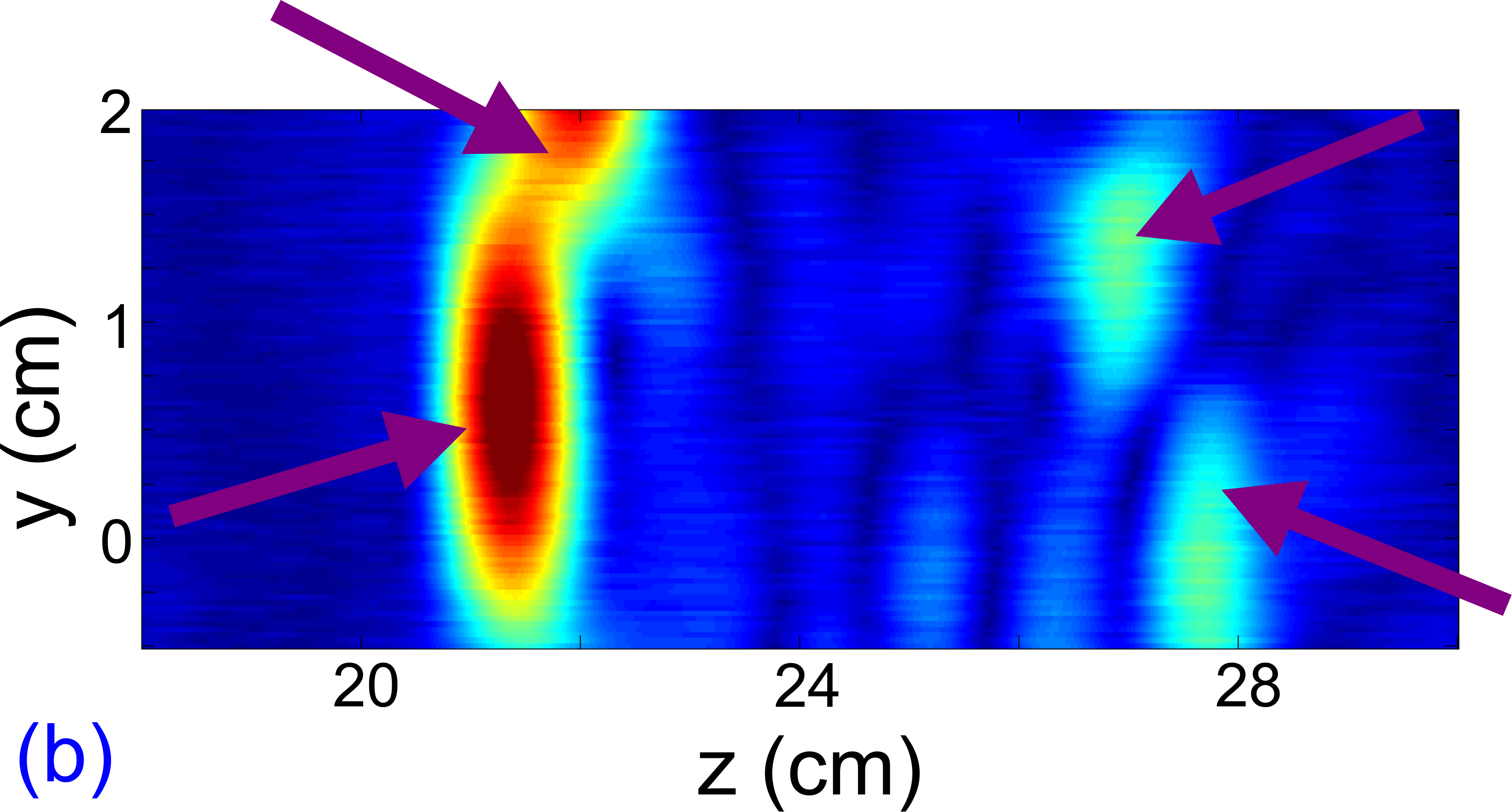}
	      \caption{\label{results_gelatin} (a) Ultrasound image of the gelatin sample. (b) Electrical impedance image of the gelatin sample. Arrows are indicating the interfaces shown on the photograph.}
		\end{center}
\end{figure}

The figure (\ref{results_gelatin}) shows respectively the ultrasound image and the electrical impedance image of the gelatin phantom. The front and back interfaces are equally visible on both images. The relative amplitude of interface signal is however different, mostly due to magnetic field inhomogeneity.

\begin{figure}[ht]
		\begin{center}
	   		\includegraphics[width=.9\linewidth]{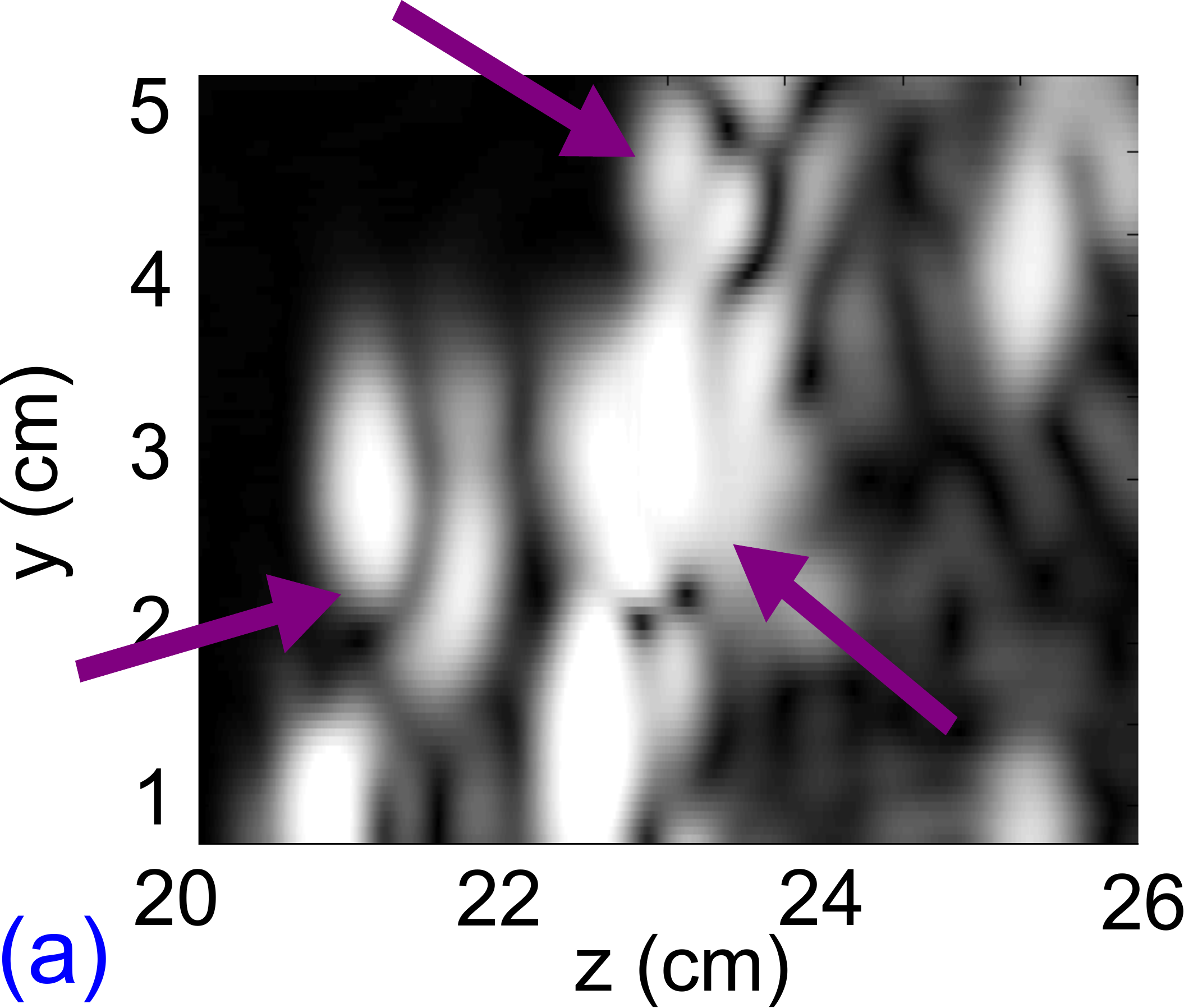}
		\end{center}
		\begin{center}
	   		\includegraphics[width=.9\linewidth]{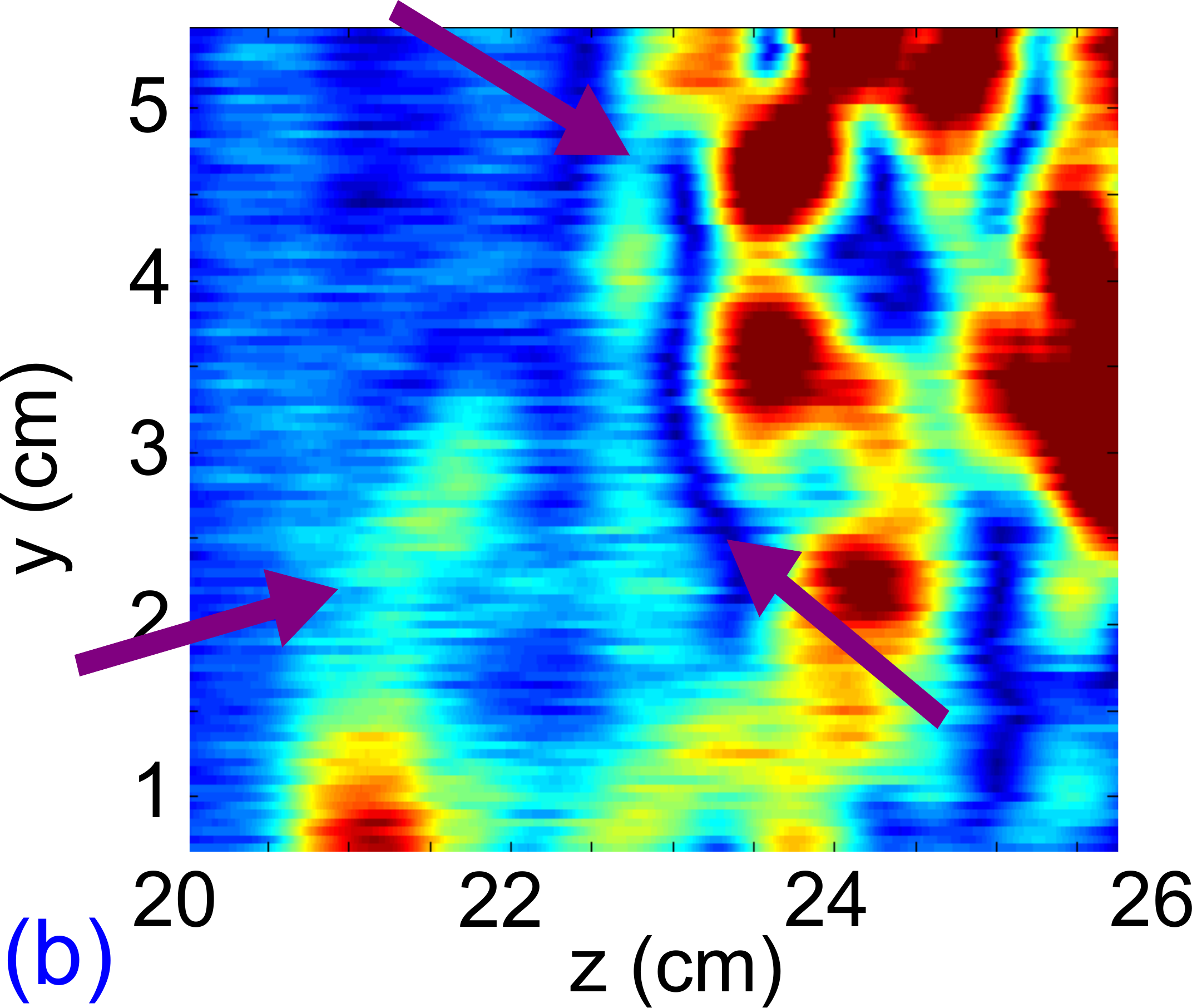}
	   		\caption{\label{results_beef} (a) Ultrasound image of the beef sample. (b) Electrical impedance image of the beef sample. Arrows are indicating the interfaces shown in the photograph.}
		\end{center}
\end{figure}

The figure (\ref{results_beef}) shows respectively the ultrasound image and the electrical impedance image of the beef sample. The front interface can be seen on both images, although the lower one is less visible.

\section{Discussion and conclusion}
A method to observe electrical conductivity gradients by combining ultrasound and magnetic field is tested experimentally.

The tested transducer is a standard single element ultrasonic therapy transducer. The magnetic field is created by a permanent magnet with sufficient air gap to insert a tissue sample. The signal has been shown to be little influenced by the electrodes positions \cite{montalibet2002these}. The oil bath prevents any electrical contact between the sample and the transducer; in practical applications it  could easily be placed around the transducer rather than around the patient. The experiment showed images of quality comparable to the one of the ultrasound images taken in the same conditions. The combination of ultrasonography and Lorentz Force Electrical Impedance Tomography with ultrasonically-induced Lorentz force can be made easily with the same material.  Elements like fat layers which are hardly visible here in the ultrasonic image can be observed in the Lorentz Force Electrical Impedance Tomography image. The technique has potentially the spatial resolution of the ultrasound wavelength \cite{islam1988}, allowing the observation of small inhomogeneities and can help thus revealing pathologies like cancer by detecting tumorous tissues where other techniques fail to do so. 

Image quality could be improved with a higher ultrasound frequency and a thinner beam. The compatibility of ultrasound imaging with MRI \cite{haigron2010} shows that it is possible to use much stronger magnetic fields, which would increase the intensity of the induced electrical current, and thus the signal to noise ratio. Moreover, even if the magnetic field homogeneity is not as critical in this technique as in MRI \cite{montalibet2002scanning}, a more homogeneous magnetic field would provide sharper images of the interfaces.
The technique can also be used in a reverse mode \cite{wen1998}, with an electrical current applied in a tissue submitted to a magnetic field, which leads to ultrasound wave \cite{xu2005} and was recently applied to human tissue ex-vivo \cite{hu2011}. It is nevertheless hard to say which of these techniques would be the most useful for biomedical imaging \cite{roth2011}.

\section{Acknowledgments}
The authors would like to thank Amalric Montalibet for its pioneer work on the subject, Rémi Souchon for his advices and Alexandre Petit for theory suggestions.No conflict of interest is declared.

\bibliographystyle{plain}
\bibliography{biblio}
\end{document}